\documentclass[
    ,final            
    ,sort&compress
  ]
  {aipproc}

\layoutstyle{8x11double}

\begin{document}

\title[Physical Properties of RRd stars]{Physical properties of double-mode RR~Lyrae stars based on pulsation and evolution models}




\author{I.~D\'ek\'any}{address={Konkoly Observatory, PO Box 67, H-1525, Budapest, Hungary}}

%

\begin{abstract}
We determine the fundamental stellar parameters of altogether 20 
double-mode RR~Lyrae (RRd) stars from the Galactic Field and the Large 
Magellanic Cloud, following our approach employed on the field 
RRd star BS~Comae \citep{bscompap}. The stars 
were selected to cover wide ranges of periods and period ratios, 
implying diverse stellar parameters. From the possible observed 
quantities we use only the periods and determine stellar parameter 
combinations that satisfy both current helium-burning evolutionary 
models and grids of linear non-adiabatic purely radiative pulsational 
models. Thus, the periods of an object determine a sequence of 
solutions for its mass, luminosity, effective temperature and 
metallicity, parametrized by the time elapsed from the zero age 
horizontal branch. The derived sets of solutions yield various 
important theoretical relations between the physical parameters 
of the stars which are, in the case of some parameter combinations, 
nearly independent of the age. We get very tight simple linear 
relations between $\log{P_0}$ and $\log{R}$, $\log{\rho}$, $\log{g}$ 
and the Wesenheit index $W(B-V)$. This latter period-luminosity-color 
relation is in good agreement with the one derived on empirical 
basis \cite{kw01} and calibrated by Baade-Wesselink results 
\cite{kovacs2003}.
\end{abstract}

\maketitle


\section{Introduction}

Double-mode RR~Lyrae (RRd) stars pulsate simultaneously and stably in the
first two radial 
modes with a period ratio close to 0.75. Their two precisely observable 
periods make them a particularly important subtype of RR~Lyrae stars 
by supplying an additional constraint on their structure. Eventually, 
the double-mode nature offers the possibility of determining the 
fundamental parameters of these stars from radial pulsation models. 
The solid theory of linear radial pulsation has long been used 
for this purpose by fitting models simultaneously to the periods of 
both pulsation modes. By measuring or giving an estimate on the 
chemical composition and effective temperature of the star we can 
derive its mass and luminosity. The consistency of these values can be 
confirmed by independently obtained distances 
\citep[see, e.g.,][]{kovacs_lmcrrd}.

Nonlinear hydrodynamical modeling of double-mode pulsation has also 
became possible in the past decade \citep[][]{feuch,szkb_nlrrd}. Sustained 
double-mode pulsation can be reproduced in a consistent range of 
physical parameters, a star-by-star modeling, however, is still not available.

For a fully consistent RRd model it is important to 
take horizontal branch (HB) stellar evolution also into account. 
In a novel study of this endeavor, to involve evolutionary constraints 
in explaining the observed period distribution of RRd stars 
of the Large Magellanic Cloud (LMC) by means of physical parameter ranges, 
\cite{pop2000} calculated pulsation model sequences along theoretical 
tracks of HB evolution to compare the theoretical period 
distribution with the observed one. In the present study we follow a 
different approach by taking the observed periods as an input to our 
calculations and search for a set of physical parameters that 
satisfy the combined parameter space of pulsational and evolutionary models. 
Thus obtaining strongly constrained values for the fundamental parameters 
of a representative sample of RRd stars, we can probe the full parameter 
ranges they populate, as well as use them to deduce theoretical relations 
among their physical parameters.

\section{Method and data}

Our goal is to determine the fundamental parameters of an RRd 
star solely from its periods by restricting the parameter space 
of pulsation models by the constraints of evolutionary models. 
The basic assumption behind this approach is that there is a 
good level of consistency between the two kinds of models. We 
obtain the fundamental physical parameters of a star by the 
same procedure as described in \citep{bscompap}, and hereby 
we only briefly summarize the essentials of the method. 

On the one hand we calculate a grid of linear pulsation models fitting 
the observed pair of periods. These models give stellar mass and 
luminosity for specified effective temperature and metallicity 
values. On the other hand, isochrones of Helium-burning models 
also determine mass and luminosity as a function of effective 
temperature, metal content, and age. To obtain the set of parameters 
that satisfy both models we have to calculate the intersection 
of these two restricted parameter-subspaces (formally vector-vector 
functions that we sample on a grid). By fixing the periods to the 
observed ones, evolutionary models have one variable in surplus, 
the age, ie. the time elapsed from the zero age horizontal branch 
(ZAHB), which will be a free parameter of the solution.

We used the canonical HB evolution models of \citep{pietr2004} 
with solar-scaled heavy element mixture. These models have a 
fine mass and time spacing, suitable for the detailed 
investigation of the instability strip. We obtained isochrones by 
interpolating the tracks to a uniform mass and time base. We generated 
linear, non-adiabatic (LNA), fully radiative pulsation models. 
The pulsation model grid was calculated for the same chemical compositions 
as those of the evolutionary tracks, and with a fine temperature step of 
$50$\,K. The deep and dense sampling of the 
stellar envelopes resulted in an accuracy of $\sim10^{-5}$ in the 
period fit. See \citep{bscompap} for further details on the model 
construction. 

\begin{figure}
  \includegraphics[width=\columnwidth]{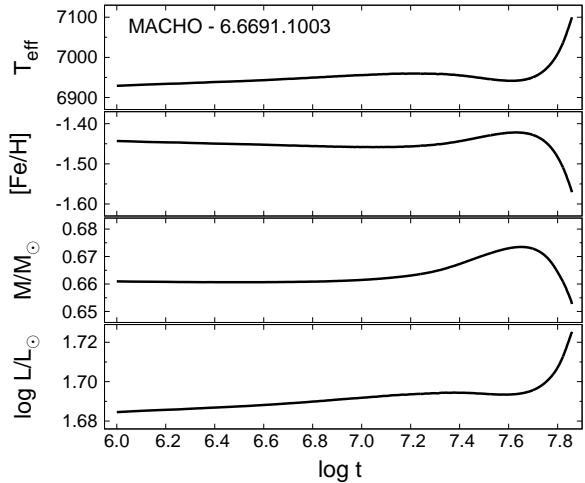}
  \caption{An example for the combined pulsational and evolutionary 
           solution sequences calculated for an RRd star in the LMC. 
           Age dependence of the various physical parameters are shown 
           up to $\log{t}=7.86~(\approx 72.5\,{\rm Myr})$.}
  \label{example}
\end{figure}

As a result of the method, for a pair of observed periods we 
obtain a 1-D solution sequence giving the effective temperature, 
metallicity, stellar mass, and luminosity as a function of the 
time parameter. Thus, each point of a sequence determine 
combinations of physical parameters corresponding to different 
stages of evolutionary tracks of different masses and metallicities. 
We also calculate synthetic colors from the parameter 
sequences using the stellar atmosphere models of \citep{atlas}. 
Fig.~\ref{example} gives an example solution for the various physical 
parameters. We note that in the absence of any additional reliable 
observational constraint, the models themselves alone do 
not pose any limit on the time parameter. However, we can still 
put rough boundaries on the solution sequences by keeping them 
within reasonable temperature ranges or omitting more advanced 
evolutionary stages that are more unlikely to be observed due to fast 
parameter changes. (Also, solutions that would require extremely low 
metallicities outside the evolutionary model grid can technically 
confine the possible ages.) Thus, in most cases we stop scanning in age 
between $\log{t}=7.85$ and $7.9$.

\begin{figure}
  \includegraphics[width=\columnwidth]{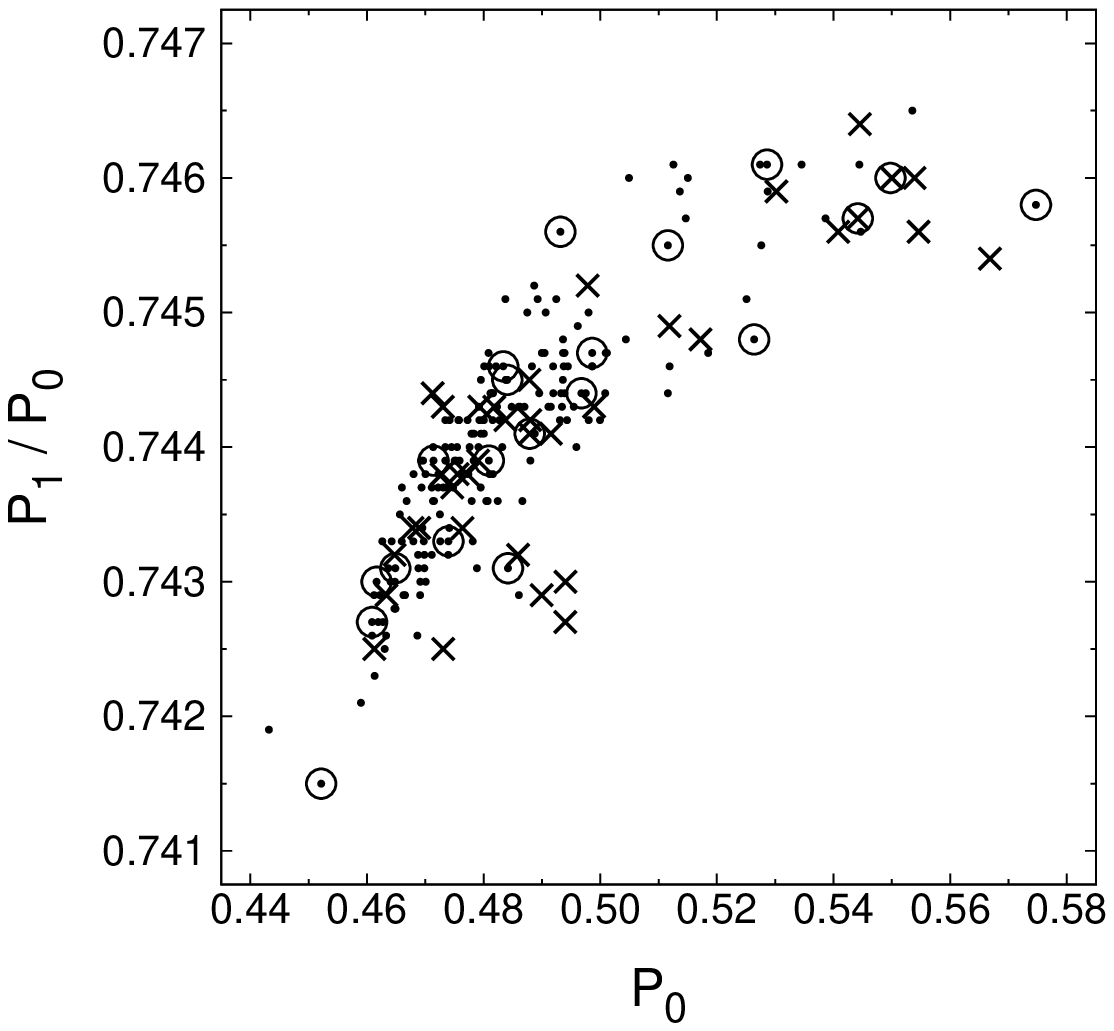}
  \caption{Petersen diagram showing RRd stars in the LMC, taken from 
           the MACHO catalog \citep[][dots]{kovacs_lmcrrd}, together 
           with all known Galactic Field RRd stars taken from the 
           literature \citep[][crosses]{wils2006zgru,oaster,bernhard2006,
           dekany2007,pwils_sdss,khruslov2007,kryachko2008,hajdu2009}. 
           Stars analyzed is this study are denoted by circles.}
  \label{petersen}
\end{figure}

Figure~\ref{petersen} shows the Petersen diagram of all known Galactic 
Field RRd stars and those found in the LMC by the MACHO project. 
The ranges of the $P_1/P_0$ vs. $P_0$ distribution populated by 
these stars encompass the vast majority of all known double-mode 
RR~Lyrae stars in any stellar system. We selected a sub-sample of 20 
stars (3 from the Field: AQ~Leo, BS~Com, and CU~Com, and 17 more from 
the LMC) to survey full ranges of physical parameters, as implied by 
the periods and period ratios. Sequences for fundamental physical 
parameters were obtained for each of the stars in the way discussed 
above. We present the analysis of the derived sets of solutions in 
the following section. 

\section{Results}

Based on the Petersen diagram, the physical parameters obtained 
for the 20 objects investigated in this work are representative 
for all known RRd stars (and their luminosities, presumably, for a 
dominant fraction of single-mode RR~Lyrae stars). Therefore, 
we use them to measure the fundamental parameter ranges populated 
by these objects and probe important theoretical relations defined 
by the solutions of our method. 

\begin{figure}
  \includegraphics[width=\columnwidth]{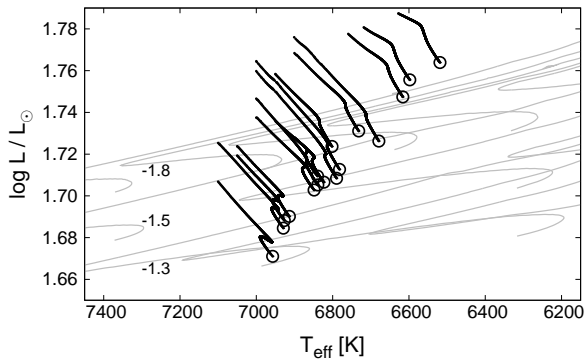}
  \caption{The loci of the RRd stars analyzed in this paper on the HRD.
           Solution sequences with free time parameter are shown as thick 
           lines. Values corresponding to zero ages are plotted by circles. 
           Helium-burning evolutionary model tracks 
           of \citep{pietr2004} are shown by thin gray lines for [Fe/H] 
           values of $-1.3$, $-1.5$, and $-1.8$ (redward parts of tracks 
           for masses lower than 0.64, 0.66, and 0.70, respectively, are not plotted for 
           more perspicuity).}
  \label{hrd}
\end{figure}

Figure~\ref{hrd} shows the distribution of the solutions on the 
Herzschprung-Russell diagram (HRD) together with various evolutionary 
tracks for reference. The expected general trend, that stars with lower metallicity are cooler 
and brighter, is clearly exhibited. The gradual topological change of 
the sequences originates from the smoothly changing sizes and shapes of 
the blueward loops in the constituting tracks, mainly as a function of 
metallicity. As it is also shown in Fig.~\ref{example}, the temperature 
dependence of a solution increases considerably for higher values of age 
(i.e., after reaching the high temperature extrema -- the `blue noses' -- 
of the constituting tracks). Therefore we would need further specific 
information on each star (i.e., on their spectral energy distributions) 
to put more significant constraints on their temperatures. However, according 
to the combined models, RRd stars populate an overall temperature range 
of at least $300$\,K, as it is clear from Fig.~\ref{hrd}. 

\begin{figure}
  \includegraphics[width=\columnwidth]{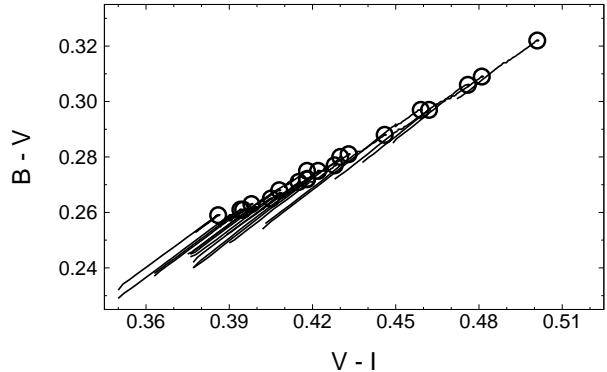}
  \caption{Two-color representation of the RRd solution sequences. 
           Notation is the same as in Fig.~\ref{hrd}.}
  \label{ci}
\end{figure}

Figure~\ref{ci} shows the solutions on the $B-V$ vs. $V-I$ two-color diagram. 
Stars are concentrated into a very narrow overall range of color index pairs. 
Here we note that by a very precisely measured pair of color indices 
one would be able to further constrain the time parameter, thus reducing 
the ranges of uncertainty in all other parameters, as it was successfully 
done in the case of BS\,Com \citep[see][]{bscompap}.

\begin{figure}
  \includegraphics[width=0.8\textwidth]{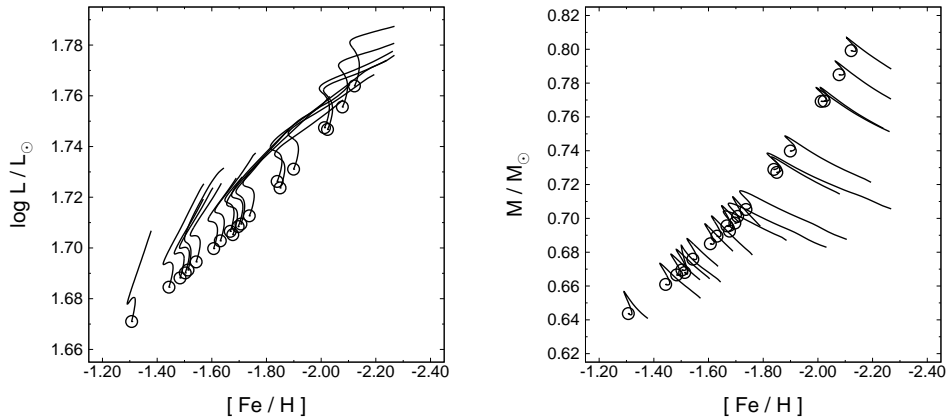}
  \caption{Luminosity-metallicity and mass-metallicity plots 
           of the RRd stars studied in this paper. For notation see 
           Fig.~\ref{hrd}.}
  \label{ml}
\end{figure}

In Fig.~\ref{ml} we examine the luminosity and mass as 
functions of the metallicity. Values corresponding to the ZAHB states 
show a tight and, apparently, somewhat non-linear correlation with 
[Fe/H] in both cases. Allowing later stages of HB evolution complicates 
the topology, but the models set minimal luminosity and maximal mass 
values at any specified metal content. 
Accordingly, for specified mass the models give a minimal luminosity, 
and, vice-verse, for a fixed luminosity, they put a constraint on the 
maximum value of the mass. Since we selected our stars to cover nearly 
the full RRd period ranges of the LMC (and by this selection, that of 
almost all known RRd stars as well), the solutions inherently yield the 
full theoretical [Fe/H], mass, and $M_V$ ranges spanned by the constituting objects. 
Although the individual sequences suffer from an uncertainty of roughly 
$0.1-0.3$\,dex in metallicity (depending on the allowed range of the time parameter), 
the upper limit for the overall [Fe/H] range turns out to be 
$\sim 0.9$\,dex with extrema of approximately $-1.3$ and $-2.2$. 
We note that, as implied by the Petersen diagram (ie., the short period 
end is more populated in the case of the LMC), the majority of the stars 
are in the upper part of this metallicity range. By direct spectrophotometric 
measurements of 6 RRd stars in the LMC, \citep{bragaglia2001} found a wider 
metallicity range of ($-2.28,-1.09$). Besides the individual errors of the 
few measurements and the sensitivity of the $\Delta S$ method to different 
calibrations, this moderate difference may result from the possible non-solar 
heavy element mixture of these stars (we recall that throughout this paper 
solar-scaled mixtures of heavy elements are used).

The stellar masses corresponding to our metallicity domain 
span in a large $0.17\,{\rm M}_\odot$ range from $0.64\,{\rm M}_\odot$ 
to $0.81\,{\rm M}_\odot$. As for absolute magnitudes, the luminosity 
interval permitted by the solutions on (see Fig.~\ref{ml}) 
can be translated into an $M_V$ range of $(0.46,0.70)$ which, as we 
will see later, is in a good agreement with Baade-Wesselink results. 

\begin{figure}
  \includegraphics[width=\textwidth]{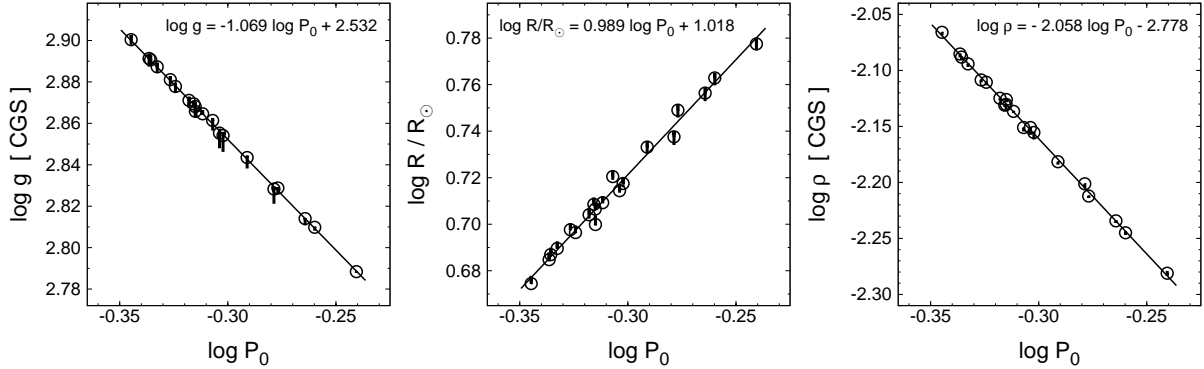}
  \caption{Surface gravitational acceleration, stellar radius, and stellar 
           density as a function of the period of the fundamental mode. 
           Notation is the same as in Fig.~\ref{hrd}. Lines show linear fits 
           to the zero age solutions, yielding the formulae given in each panel.}
  \label{p0corrs}
\end{figure}

We also computed $\log{g}$, $\log{R}$, and $\log{\rho}$ values for all 
solution sequences. These parameters are highly degenerate, i.e., very 
insensitive to the time parameter. They also show very tight, linear 
correlations with the fundamental period. 
Figure~\ref{p0corrs} shows these simple \emph{theoretical} relations provided 
by our results. It is remarkable how tiny individual parameter ranges are 
allowed by the solutions. Thus, for the sake of simplicity, linear formulae 
given on the same figure were obtained by fitting to the zero age solutions 
only. 

\begin{figure}
  \includegraphics[width=\columnwidth]{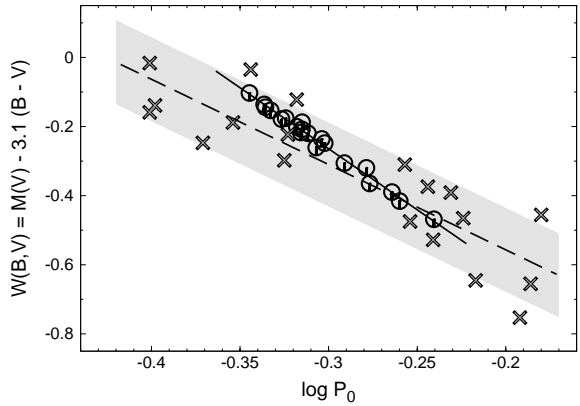}
  \caption{PLC relation for the RRd sample of this study are shown with 
           the same notation as in Fig.~\ref{hrd}. Crosses denote BW results of 
           \citep{kovacs2003}. Solid line shows the linear fit to the RRd data. 
           PLC relation of \citep{kw01} with zero-point fixed by the BW sample 
           is delineated by a dashed line, with a shaded area illustrating 
           the $3\,\sigma$ ranges of its calibrating sample. 
           }
  \label{plc}
\end{figure}

One of the tightest empirical relations for RR~Lyrae stars is the 
period-luminosity-color (PLC) relation, or in other words the period -- 
reddening-free Wesenheit-index relation. 
We can directly compare it with the one derived from the synthetic $M_V$ 
and color index values of RRd stars calculated above. 
Figure~\ref{plc} shows the theoretical PLC relation for $B$ 
and $V$ colors stretched by our RRd model results. 
In this representation of the solutions we notice a vast insensitivity again 
to evolution effects. 
For comparison we show the empirical relation derived from a large number of 
RRab stars from Galactic globular clusters (GCs) by \citep{kw01}, together 
with the $3\,\sigma$ range of the calibration sample. The zero-point of this relation 
was set by the Baade-Wesselink (BW) results for Galactic Field RRab stars 
(also shown in Fig.~\ref{plc}) as given by \citep{kovacs2003}. The zero 
points of the two relations show a fine agreement. Although the RRd sample 
clearly defines a significantly steeper slope, it does not exceed the $3\,\sigma$ 
ranges of the GC sample. We note that the situation is very similar in the case 
of the $V$ and $I$ colors. 

The reason for the slope difference is not yet clear and may be caused by a 
combination of different effects. Since the PLC relation is governed by 
two parameters (i.e., metallicity and effective temperature) out of four 
of the pulsation equation, it has an intrinsic scatter within which the 
subset of RRd stars could follow their own, tighter relation, determined by 
the specific stellar parameter subspace they populate. 

We can also speculate on the following. 
In the derivation of the empirical PLC relations by \citep{kw01}, no direct 
calibration (i.e., using independently measured absolute magnitudes) was made. 
Instead, the relative dereddened distance moduli of the constituting globular 
clusters were obtained as output parameters of the least-squares minimization 
procedure, together with the regression coefficients. Thus, a uniform slope 
was obtained for all stars, nicely fitting the subsamples of the individual 
clusters, and the zero-point differences were interpreted as the result of the 
relative distances. Later \citep{soszynski2003} found a somewhat steeper 
$W(V,I)-\log{P_0}$ relation for LMC RRab stars than that of \citep{kw01}, but 
the scatter was much higher. The HB stars of a GC have generally little if any 
metallicity dispersion. Therefore, if an additional slight, intrinsic metallicity 
dependence of the Wesenheit index does rather manifest in zero point than 
in slope differences between subsamples of stars with constant metallicity, its 
effect could merge with that of the relative distances and remained hidden in 
the above case. Furthermore, if this is so, then our RRd sample will yield a 
steeper relation because it scans a considerable, nearly 1\,{\rm dex} range 
in [Fe/H], which, in this case, is more correlated with $\log{P_0}$. 
However, we would need very precise independent relative distance determinations 
for the calibrating clusters to give a more secure assessment of this problem. 

\section{Conclusions}

We presented a method and its applications for stellar parameter determination 
of double-mode RR~Lyrae stars. The method combines theoretical model results of linear 
stellar pulsation and canonical HB stellar evolution. Based on the periods of an 
object, the joint constraints of the models allow sequences of fundamental stellar 
parameters with age as a free parameter. Additional observables like color index 
or metallicity can be used to put further constraints on the solutions. By using 
a representative sample of objects to probe the observed period ranges of the 
known RRd's with our method, we obtained the theoretical ranges of physical parameters 
populated by these objects. We also extracted important theoretical relations 
from the derived sets of stellar parameters among which we found nearly age-independent 
ones. We derived a very tight linear PLC relation with a magnitude zero-point being 
in a fine agreement with that of Baade-Wesselink results. We found a slope difference, 
which might be explained by a hitherto hidden metallicity dependence of the relation.


\begin{theacknowledgments}
The author is very grateful to G\'eza Kov\'acs for the many fruitful 
discussions regarding double-mode pulsation and empirical relations 
of RR~Lyrae stars. Financial support of the Hungarian Research Foundation 
(OTKA) grant K-60750, as well as from the ELTE Doctoral School of Physics, 
and from the organizers of the 19th pulsation meeting of the ``Los Alamos 
Series'' are highly appreciated. 
\end{theacknowledgments}

\bibliographystyle{aipproc}   

\bibliography{dekany}

\end{document}